# On the force exerted on a non-spherical dust grain from homogeneous, stationary, non-magnetized plasma

S. I. Krasheninnikov and R. D. Smirnov

*University California San Diego, 9500 Gilman Dr., La Jolla, 92093 CA, USA*

**Abstract**

It is shown that stationary non-spherical dust grain immersed into stationary non-magnetized plasma can experience a force caused by the grain-plasma interactions.

The dynamics of dust (micro-particles of an effective radius from nanometers to tens of microns) often appear to be an important ingredient in different plasma-related phenomena in astrophysics, physics of the solar system, and laboratory experiments [1-6]. Therefore, there are many theoretical papers devoted to the studies of the forces and torques exerted on the dust grain in the course of grain-plasma interactions (e.g. see [5-10] and the references therein). In most cases, these studies assumed spherical grain shapes, and such forces and torques were associated with an electric field, relative grain and plasma velocities, plasma temperature gradients, shear of plasma flow, gyro-motion of plasma electrons and ions, etc.

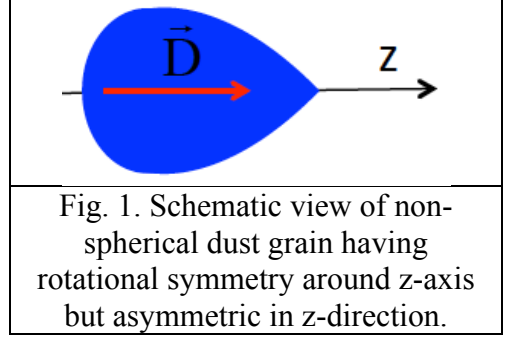

Fig. 1. Schematic view of non-spherical dust grain having rotational symmetry around z-axis but asymmetric in z-direction.

However, in practice, the shapes of natural dust grains are far from being spherical (e.g. see Refs. 4, 5, 11). In this case, finding the force and torque exerted on the grain from surrounding plasma becomes a challenge. In Ref. 12, 13 it was suggested to use an approach utilizing the moments attributed to the shape of the grain combined with different coordinate transformation properties of polar- and pseudo-vectors. We notice that a similar approach was used in the studies of the dynamics of small particle embedded into fluids [14, 15]. For example, consider the grain having rotational symmetry around the z-axis, but asymmetric in the z-direction (see Fig. 1). The spatial orientation of such grain can be characterized by the unit vector $\vec{D}$. Assume now that this grain is embedded into non-magnetized plasma flowing with velocity $\vec{U}$ with respect to the grain. In case the grain is not spinning, the force, $\vec{F}$, and torque, $\vec{T}$, acting on the grain should be expressed with the vectors $\vec{D}$ and $\vec{U}$ and some scalar quantities characterizing both plasma (e.g. density, n, and temperature, T) and the dust grain (e.g. grain effective size $R_d$). Keeping in mind that $\vec{F}$ is a polar vector, whereas $\vec{T}$ is a pseudo-vector, we find

$$\vec{F} = \Phi_D \vec{D} + \Phi_U \vec{U} + \Phi_{UD} \vec{D}(\vec{U} \cdot \vec{D}), \quad \vec{T} = \hat{T}(\vec{U} \times \vec{D}), \qquad (1)$$

where $\Phi_D$, $\Phi_U$, $\Phi_{UD}$, and $\hat{T}$ are the scalars depending on both plasma and dust grain parameters. For the case where $\vec{U}$ is small ($U \ll \sqrt{T/M}$, where M is the ion mass), these scalars can be considered as independent from $\vec{U}$ [12]. Even though an exact calculation of these scalars is a difficult problem, some rather simple estimates, allowing shed light on the dynamics of non-spherical dust grains, can be made [12, 13].

As we see from Eq. (1) the formal expression for the force implies that the force, caused by grain-plasma interactions, could be exerted on the grain even by stationary plasma. For the case of fluid, it is easy to show that $\Phi_D = 0$ and such force does not exist (e.i. $\Phi_D \vec{D} = -\oint_{S_d} p d\vec{S} = 0$, where the integration goes over the entire grain surface, $S_d$).

However, recently, it was found that in the case of non-spherical grain situated in stationary magnetized plasma embedded into a magnetic field $\vec{B}$, the $\vec{j} \times \vec{B}$ force (where $\vec{j}$ is the electric current flowing through the grain) could exist [16].

We notice that in the case of inhomogeneity of the surface material, the forces due to inhomogeneity of material ablation or the accommodation of the momenta of electrons and ions impinging to the surface exist even for a spherical grain immersed into non-magnetized plasma. But, in what follows we will not consider such effects. Instead, we will present the arguments that for the case of interaction of non-spherical grain with collisionless plasma $\Phi_D$ can be finite even for no grain ablation and homogeneous surface material.

As a "proof of principals", we consider a non-spherical dust grain having rotational symmetry around the z-axis but asymmetric in the z-direction (e. g. see Fig. 1) embedded into plasma with electron temperature, $T_e$, significantly larger that the ion one, $T_i$. We also will assume that the Debye length, $\lambda_D$, is much larger than the characteristic size of the dust grain $R_d$. Finally, we will assume that the grain material is a conductor.

For $\lambda_D / R_d \gg 1$ the main mechanism of the momentum exchange between plasma particles and the grain is due to a small angle scattering. Moreover, for $T_e / T_i \gg 1$ the ion momentum exchange dominates. We note that similar conclusions hold for the case of spherical grain (e.g. see Ref. 8 and the references therein). However, for spherical grain the asymmetry of the electrostatic potential distribution around the grain, $\varphi(\vec{r})$, resulting in the asymmetry of plasma particle scattering and subsequently producing the net force, only occurs for $U \neq 0$. Contrary to that, an asymmetric equipotential surface of grain produces asymmetry of $\varphi(\vec{r})$ in the grain vicinity even for $U = 0$. As a result, the angle scattering of plasma particles approaching the grain from $z = +\infty$ and $z = -\infty$ will be different. One can illustrate it for the case where $\varphi(\vec{r})$ is described by the superposition of a charge, $Q_0$, and a dipole $\vec{\delta}_0$:

$$\varphi(\vec{r}) = \frac{Q_0}{r} + \frac{\vec{\delta}_0 \cdot \vec{r}}{r^3}, \tag{2}$$

which can be used for the proxy of the potential distribution around non-spherical grain for $r > |\vec{\delta}_0|/Q_0$. Then, assuming that $\vec{\delta}_0$ is aligned with the z-coordinate we find the following equation for the dynamics of ions moving from $z = +\infty$ and $z = -\infty$ in an electric field defined by Eq. (2)

$$M\frac{d\vec{V}}{dt} = eQ_0 \frac{\vec{r}}{r^3} + \sigma e |\vec{\delta}_0| \left( 3\frac{\vec{r}(\vec{e}_z \cdot \vec{r})}{r^5} - \frac{\vec{e}_z z}{r^3} \right), \tag{3}$$

where $\sigma = 1$ ($\sigma = -1$) for the particle moving $z = -\infty$ ($z = +\infty$). As we see from Eq. (3) the sign of $\sigma$ breaks the scattering symmetry, which causes the existence of the net force even in the case of equivalent particle fluxes coming toward the grain from $z = +\infty$ and $z = -\infty$.

In conclusion, we presented the compelling arguments for the existence of the force exerted on stationary non-spherical grain by stationary homogeneous plasma, which justifies a more general expression for the force given by Eq. (1). More detailed numerical studies of such force will be published elsewhere.

**Acknowledgements**
This work was supported by the U.S. Department of Energy, Office of Science, Office of Fusion Energy Sciences under Award No. DE-FG02-06ER54852 at UCSD.